\documentclass[10pt,conference]{IEEEtran}
\usepackage{amsmath, amssymb, amsbsy}
\usepackage{graphicx}
\usepackage{hero}

\title{Centralized Network Utility Maximization over Aggregate Flows}

\author{\IEEEauthorblockN{Riten Gupta}
\IEEEauthorblockA{UtopiaCompression Corporation\\
11150 West Olympic Blvd. \#820\\
Los Angeles, CA 90064\\
Email: riten@utopiacompression.com}
\and
\IEEEauthorblockN{Lieven Vandenberghe}
\IEEEauthorblockA{Electrical Engineering Dept.\\
University of California, Los Angeles\\
Los Angeles, CA 90095\\
Email: vandenbe@ee.ucla.edu}
\and
\IEEEauthorblockN{Mario Gerla}
\IEEEauthorblockA{Computer Science Dept.\\
University of California, Los Angeles\\
Los Angeles, CA 90095\\
Email: gerla@cs.ucla.edu}}

\def\bmu{{\boldsymbol{\mu}}}
\def\brho{{\boldsymbol{\rho}}}
\def\blambda{{\boldsymbol{\lambda}}}
\def\bsigma{{\boldsymbol{\sigma}}}
\def\b1{{\mathbf{1}}}
\newcommand{\prox}{{\mbox{\rm prox}}}

\begin{document}
\maketitle

\begin{abstract}
We study a network utility maximization (NUM) decomposition in which the set of
flow rates is grouped by source-destination
pairs. We develop theorems for both single-path and multipath cases, which
relate an arbitrary NUM problem involving all flow rates to a simpler problem
involving only the aggregate rates for each source-destination pair. The
optimal aggregate flows are then apportioned among the constituent flows of
each pair. This apportionment is simple for the
case of $\alpha$-fair utility functions. 
We also show how the decomposition can be implemented with the alternating
direction method of multipliers (ADMM) algorithm.
\end{abstract}

\section{Introduction}

The last two decades have seen a great deal of research in network utility
maximization (NUM) \cite{kelly1998rate} \cite{Palomar&Chiang:JSAC2006}
\cite{chiang2006layering}, which has cast light on traditional networking
protocols \cite{low2003duality} and has facilitated the design of promising
future protocols \cite{wei2006fast} as well. 
Most NUM researchers have focused on developing distributed solutions
to various utility maximization problems.  These distributed solutions, which
follow nicely from dual decompositions \cite{Palomar&Chiang:JSAC2006}, are
ideal for internets, in which cooperation among flow sources cannot be assumed,
and minimal communication between links and nodes is desired.  In recent years
there has been growing interest in the software defined networking (SDN)
paradigm, in which data and control planes are
separated \cite{mckeown2008openflow} \cite{mckeown2009software}.  
In this framework, certain network
functions such as flow control, congestion control, and throughput optimization
may be assigned to a central controller. Central control is feasible for closed
networks, such as in data centers \cite{benson2010network}
or communication satellite networks \cite{donner2004mpls}.

In some networks with central control, the number of flows $K$ may be much
larger than the number of source-destination pairs $N$. For example, the
Iridium satellite network employs 66 satellites and facilitates tens of
thousands of flows \cite{pratt1999operational}. 
A similar phenomenon may occur in small data centers.  In
this paper, we study a primal decomposition in which the set of flows is grouped
into flow classes, each corresponding to a source-destination pair.  
Many congestion control algorithms inherently group flows by source-destination
pair \cite{Bertsekas&Gallager:1992} and several related primal decompositions
have been studied, for example in \cite{Palomar&Chiang:JSAC2006}.  However,
because the source-destination decomposition is only applicable to centralized
control, it has received little attention. Given the recent popularity of SDN,
however, the decomposition may prove to be beneficial. To this end, we develop
a comprehensive theory of the source-destination decomposition in this paper.
(We also discuss briefly in Section \ref{sec:conclusion} a potential benefit of
this decomposition in a network with ``semi-distributed'' control.)  We derive
theorems that decompose a NUM problem with $K$ variables into one with only $N$
variables, followed by an allocation problem which apportions the aggregate
rate for each class among the class's constituent flows. In some cases,
this apportionment is simple, 
In other cases, the 
alternating direction method of multipliers (ADMM) algorithm can exploit the
decompostion numerically.

The remainder of the paper is organized as follows. In Section \ref{sec:agg} we
present the aggregate flow decomposition and the main results relating the
original NUM problem to the simpler aggregate flow problem. This analysis is
extended to the multipath case in Section \ref{sec:multipath}.  In Section
\ref{sec:algorithms} we discuss numerical algorithms,
which exploit the aggregate flow decomposition. Finally, in Section
\ref{sec:conclusion}, we conclude the paper.

\section{Optimization Over Aggregate Flows}
\label{sec:agg}

Consider a communication network with $M$ nodes and $L$ links (edges). Let $N$
be the number of source-destination pairs in use among all flows. Then 
$N \le M(M-1)$. Number the source-destination pairs $1, \dots, N$ and call 
the set
of flows in pair $i$ the $i$th {\it flow class}. Let $K_i$ be the number of
flows in class $i$. Then the total number of flows is $K = \sum_{i=1}^N
K_i$. Let $u_{i,k}$ be the rate of the $k$th flow in class $i$.  Finally,
define the $L \times N$ binary routing matrix $\bR$ as
\begin{equation*}
R_{l,i} = \left\{
\begin{tabular}{ll}
$1$, & traffic of class $i$ passes through link $l$ \\
$0$, & otherwise.
\end{tabular}
\right.
\end{equation*}
Note that all flows within a class follow the same path.
Consider the following utility maximization problem.
\begin{eqnarray}
\underset{\{u_{i,k} \}}{\text{maximize}} &~& 
\sum_{i=1}^N \sum_{k=1}^{K_i} f_{i,k} (u_{i,k})  \nonumber\\
\text{subject to} &~& \sum_{i=1}^N \sum_{k=1}^{K_i} R_{l,i} u_{i,k} \le c_l, \
l = 1, \dots, L
\label{eq:problem1}
\end{eqnarray}
where $f_{i,k}$ is a utility function for the $k$th flow in class $i$,
$c_l$ is the capacity of link $l$, and the constraints imply
that no link is overloaded. Next, let $x_i = \sum_{k=1}^{K_i} u_{i,k}$ be the 
aggregate rate of class $i$ and consider the aggregate flow utility 
maximization problem
\begin{eqnarray}
\underset{\{x_{i}\}}{\text{maximize}} &~& 
\sum_{i=1}^N f_{i} (x_{i})  \nonumber\\
\text{subject to} &~& \sum_{i=1}^N R_{l,i} x_{i} \le c_l, \
l = 1, \dots, L
\label{eq:problem2}
\end{eqnarray}
where $f_i$ is an aggregate utility function for class $i$. The domains of the
utility functions $f_{i,k}$ and aggregate utility functions $f_i$ are not
stated here, but are usually subsets of $\mathbb{R}_{+}$ as negative flow
rates are not allowed. 

\subsection{Decomposition by Supremal Convolutions}
\label{sec:sup_conv_decomp}

%
%

\begin{definitions}
Let the functions $f_1$ and $f_2$ be concave and proper on $\mathbb{R}^n$.
The supremal convolution of $f_1$ and $f_2$ is
\begin{equation*}
(f_1 \diamond f_2)(x) = \underset{\{(x_1,x_2) : x_1 + x_2 = x\}}{\sup} 
f_1(x_1) + f_2(x_2).
\end{equation*}
\end{definitions}
The supremal convolution of the concave functions $f_1$ and $f_2$ is simply the
negative infimal convolution of the convex functions $-f_1$ and $-f_2$.
By \cite[Theorem 5.4]{rockafellar1970convex}, $f_1 \diamond f_2$ is concave.
Observe that problem (\ref{eq:problem2}) is equivalent to problem 
(\ref{eq:problem1}) when each $f_i$ is the $K_i$-fold supremal
convolution
\begin{equation*}
f_i(x_i) = 
\underset{\substack{  \{u_{i,1}, \dots, u_{i,K_i}\} :\\ \sum_k u_{i,k} = x_i }}
{\sup}
\sum_k f_{i,k} (u_{i,k}) 
= (f_{i,1} \diamond \dots \diamond f_{i,K_i})(x_i)
\label{eq:supremal}
\end{equation*}
%
%
and for each flow class $i$, the optimal subflow rates
solve 
\begin{equation*}
\underset{\{u_{i,k} \}}{\text{maximize}} 
\sum_{k=1}^{K_i} f_{i,k} (u_{i,k}), \quad
\text{subject to}  \sum_{k=1}^{K_i} u_{i,k} = x_i^*
\end{equation*}
where $x_i^*$ is the solution of problem (\ref{eq:problem2}) 
(provided it exists) with $f_i$ defined as above.
Thus, when the utilities are concave and proper, 
problem (\ref{eq:problem1}) can be decomposed into an
aggregate optimization and $N$ optimizations
over the subflows as long as the supremal convolutions can be
calculated.
This decomposition lends itself to parallel implementations, as the 
$N$ subproblems are independent.

Let $f^*(y) = \inf_x (xy - f(x))$ denote the concave
Fenchel conjugate. From \cite[Theorem 16.4]{rockafellar1970convex}, 
the conjugate supremal convolution is 
$(f_{i,1} \diamond \dots \diamond f_{i,K_i})^* = \sum_k f_{i,k}^*$.
Thus the concave closure of the supremal convolution is $(\sum_k f_{i,k}^*)^*$.
By \cite[Corollary 20.1.1]{rockafellar1970convex}, if the $f_{i,k}$'s are
closed
and $\cap_k \text{relint} (\text{dom} f_{i,k}^* ) \ne \emptyset$,
then $f_{i,1} \diamond \dots \diamond f_{i,K_i}$ is closed, so
\begin{equation}
f_i = f_{i,1} \diamond \dots \diamond f_{i,K_i} = (\sum_k f_{i,k}^* )^*.
\label{eq:supremal_conj}  
\end{equation}

\subsection{Decomposition with Functions of Legendre Type}

\begin{definitions}
A pair $(f, \calD)$ is of Legendre type if $\calD$ is a nonempty 
open convex set, $f$ is a strictly concave differentiable function on $\calD$
and $\lim_{n \to \infty} \|\nabla f(x_n)\| = +\infty$ for any sequence
$\{x_n\}$ in $\calD$ converging to a boundary point of $\calD$.
\label{def:legendre}
\end{definitions}

Although the Legendre type property applies to a pair $(f, \calD)$, we
will refer to a function $f$ as being of Legendre type when 
$(f, \text{int}( \text{dom} f))$ is of Legendre type. Note that when 
$\text{dom} f = \mathbb{R}_{++}$, the last condition in Definition 
\ref{def:legendre} is equivalent to $\lim_{x \downarrow 0}f'(x) = +\infty$.

The (concave) Legendre conjugate (or Legendre transform)
\cite{rockafellar1970convex} \cite{boyd2004convex} of a pair $(f, \calD)$, 
where $\calD \subset \mathbb{R}$ is open and $f$ is differentiable on $\calD$,
is the pair
$(g, \calE)$ where $g(y) = y f'^{-1}(y)  - f(f'^{-1}(y))$
and $\calE$ is the image of $\calD$ under $f'$.

From \cite[Theorem 26.5]{rockafellar1970convex}, 
if $f_{i,k}$ is closed, $\calD = \text{int}(\text{dom} f_{i,k})$,
and $\calD^* = \text{int}(\text{dom} f_{i,k}^*)$, then 
$(f_{i,k}, \calD)$ is of Legendre type if and only if
$(f^*_{i,k}, \calD^*)$ is of Legendre type.
When these pairs are of Legendre type, $(f_{i,k}^*,\calD^*)$ is the Legendre
conjugate of $(f_{i,k}, \calD)$, which is the Legendre conjugate
of $(f_{i,k}^*,\calD^*)$, so conjugation is involutory:
$(f^{**}_{i,k}, \calD^{**}) = (f_{i,k}, \calD)$, and
\begin{equation}
(f_{i,k}^{*})' = f_{i,k}'^{-1}.
\label{eq:legendre_relation}
\end{equation}
Note that if the $f_{i,k}$'s are closed and Legendre type, with 
domain $\calD$ and $\cap_k \text{relint} (\text{dom} f_{i,k}^* ) \ne \emptyset$,
then $\sum_k f^*_{i,k}$ is Legendre type
(and therefore differentiable), and  the supremal convolution
$(\sum_k f^*_{i,k})^*$ is closed and Legendre type.

Let the $f_{i,k}$'s have domain $\calD = \mathbb{R}_{++}$. 
Since there are no equality constraints in problems (\ref{eq:problem1}) and
(\ref{eq:problem2}) and the inequality constraints are all affine, Slater's
condition guarantees strong duality for each problem as long as a feasible
point exists in the relative interior of the problem domain
\cite[Sec. 5.2.3]{boyd2004convex}, which is $\mathbb{R}_{++}^K$ for problem
(\ref{eq:problem1}) and $\mathbb{R}_{++}^N$ for (\ref{eq:problem2}).  Clearly,
setting all optimization variables to a small $\epsilon > 0$ yields such a
point, so strong duality holds for both problems. 
With Legendre type functions, problem
(\ref{eq:problem1}) is strictly concave with convex feasible region and has a
unique solution. Thus there is a unique primal-dual optimal pair satisfying
the Karush-Kuhn-Tucker (KKT) conditions for problem (\ref{eq:problem1}) with
Legendre-type utility functions.

\begin{theorems}
Let the functions $f_{i,k}$ be closed, concave, and Legendre type with domain
$\mathbb{R}_{++}$ and 
$\cap_k \text{relint} (\text{dom} f_{i,k}^* ) \ne \emptyset$.
For each $i,k$, let $g_{i,k}=f_{i,k}^*$,
$g_i = \sum_k g_{i,k}$, and $f_i = g_i^*$.
Let $\{x_i^*\}$ be a primal solution to problem (\ref{eq:problem2}) with
this definition of $\{f_i\}$. Then (\ref{eq:problem1}) has unique primal
solution
\begin{equation}
u_{i,k}^* = g'_{i,k} ( f_i'(x_i^*)), \ \forall i,k
\label{eq:sp_usolution}
\end{equation}
and the corresponding dual solutions of (\ref{eq:problem1}) 
and (\ref{eq:problem2}) are equal.
\label{theorem:legendre_sp}
Note that $f_i = f_{i,1} \diamond \dots \diamond f_{i,K_i}$.
\end{theorems}

{\it Proof.} First note that $f_i$ and $g_{i,k}$ are Legendre type
and therefore differentiable. Let $h_{i,k} = g'_{i,k}$ for each $i,k$.
From (\ref{eq:legendre_relation})
we have $h_{i,k} = f'^{-1}_{i,k}$. 
The Lagrangian for problem (\ref{eq:problem1}) is
\begin{equation*}
\calL_1(\bu, \brho) = \sum_{i} \sum_{k} f_{i,k} (u_{i,k}) - 
\sum_{l=1}^L \rho_l ( \sum_{i}  \sum_{k} R_{l,i} u_{i,k} - c_l).
\end{equation*}
%
The KKT
sufficient conditions for optimality of (\ref{eq:problem1}) are
\begin{eqnarray}
\label{eq:kkt1.2}
\sum_i R_{l,i} \sum_k u_{i,k} &\le& c_l, \ \forall l \\
\label{eq:kkt1.3}
\brho &\ge& \mathbf{0} \\
\label{eq:kkt1.4}
\rho_l (\sum_i R_{l,i} \sum_k u_{i,k} - c_l) &=& 0, \ \forall l \\
\label{eq:kkt1.5}
u_{i,k} &=& h_{i,k} (\brho^T \bfr_i), \ \forall i,k
\end{eqnarray}
where $\bfr_i$ is the $i$th column of $\bR$. Condition (\ref{eq:kkt1.5})
is equivalent to $\partial \calL_1/\partial u_{i,k} = 0$.
Now, let 
\begin{equation*}
h_i = \sum_k h_{i,k} = \sum_k g'_{i,k} = g_i'
\end{equation*}
for each $i$ and consider problem (\ref{eq:problem2}) with $f_i=g_i^*$
Since $f_i$ is Legendre type, it is strictly concave and thus $\{x_i^*\}$ is
the unique solution to problem (\ref{eq:problem2}).  Since, in addition, $f_i$
is closed, we can use (\ref{eq:legendre_relation}) to get $h_i = f_i'^{-1}$.
The Lagrangian is
\begin{equation*}
\calL_2(\bx, \blambda) = \sum_i f_{i} (x_i) - 
\sum_l \lambda_l ( \sum_{i} R_{l,i} x_{i} - c_l).
\end{equation*}
The KKT conditions for problem (\ref{eq:problem2}) are thus
\begin{eqnarray}
\label{eq:kkt2.2}
\sum_i R_{l,i} x_{i} &\le& c_l, \ \forall l \nonumber\\
\label{eq:kkt2.3}
\blambda &\ge& \mathbf{0} \nonumber\\
\label{eq:kkt2.4}
\lambda_l (\sum_i R_{l,i} x_{i} - c_l) &=& 0, \ \forall l \nonumber\\
\label{eq:kkt2.5}
x_{i} &=& h_{i} (\blambda^T \bfr_i), \ \forall i.
\end{eqnarray}
Next, let $(\bx^*,\blambda^*)$ be the primal-dual solution for
problem (\ref{eq:problem2}) and set $\brho = \blambda^*$.
Then condition (\ref{eq:kkt1.3}) is immediately satisfied. Next let
\begin{equation*}
u_{i,k} = h_{i,k} (\blambda^{*T} \bfr_i) = h_{i,k} (f_i' (x_i^*)), \ \forall i,k.
\end{equation*}
%
%
%
Then condition (\ref{eq:kkt1.5}) is satisfied and using 
$h_i = f_i'^{-1} = \sum_k h_{i,k}$, and (\ref{eq:kkt2.5}), we have
%
\begin{equation*}
\sum_k u_{i,k} = \sum_k h_{i,k} (\blambda^{*T} \bfr_i) = 
h_i(\blambda^{*T} \bfr_i) = x^*_i, \
\forall i
\end{equation*}
which ensures $\sum_k u_{i,k}^* = x_i^*$, and therefore conditions 
(\ref{eq:kkt1.2}) and (\ref{eq:kkt1.4}) are satisfied. Finally,
since the image of $h_{i,k}$ is $\mathbb{R}_{++}$, 
$u_{i,k} > 0$ for each $i,k$.
\qed

\subsection{Examples}

Here we apply the aggregate flow decomposition to some example utility 
maximization problems. 
Theorem \ref{theorem:legendre_sp} can be applied to 
utility functions belonging to the family of $\alpha$-fair
functions 
\cite{srikant2013communication}, while the decomposition using supremal
convolutions must be used for more general problems.

\subsubsection{Weighted Logarithm Utilities}
\label{sec:weighted_log_example}

Let $f_{i,k} (u_{i,k}) = w_{i,k} \log u_{i,k}$
with $w_{i,k} \ge 0$, so that the overall utility is a sum of weighted 
logarithms of individual flows.  These functions belong to the class of
$\alpha$-fair utilities with $\alpha=0$ and are appealing as they
yield proportionally fair rate allocations \cite{srikant2013communication}.
They are also clearly
Legendre type so Theorem \ref{theorem:legendre_sp} can be used. 
The Legendre conjugates can be calculated using
(\ref{eq:legendre_relation}). We have
$f_{i,k}'(u) = w_{i,k} / u$, $g'_{i,k}(v) = w_{i,k} /v$,
and $g'_i(v) = \sum_k w_{i,k} / v$. 
Next $f_i'(x) = g_i'^{-1}(x) = \sum_k w_{i,k} / x$ and
$f_i(x) = \sum_k w_{i,k} \log x$. 
Finally, using (\ref{eq:sp_usolution})
\begin{equation}
u_{i,k}^* = \frac{w_{i,k}} {\sum_{k'} w_{i,k'}}  x_i^*.
\label{eq:example1_subflow}
\end{equation}
Note that all of the above functions have domain $\mathbb{R}_{++}$.
From (\ref{eq:example1_subflow}),
the optimized aggregate flows should be apportioned to
the subflows in proportion to their weights. Note that the utility function
of problem (\ref{eq:problem2}) is also a sum of weighted logarithms, where the
$i$th weight is the sum weight of the $i$th class.

The weighted logarithm case can also be proven using proportional 
fairness \cite{kelly1997charging}. Let
$\{x_i^*\}$ be the solution to problem (\ref{eq:problem2}) with $f_i(x_i) =
w_i \log x_i$ and let the subflow rates be 
\begin{equation}
u_{i,k} =  \frac{w_{i,k}} { w_i} x_i^*
\label{eq:subflow_alloc}
\end{equation}
where $\{w_{i,k}\}$ are any non-negative weights such that 
$\sum_k w_{i,k} =w_i$.  From \cite{kelly1997charging}, the unique solution to
(\ref{eq:problem2}) $\{x_i^*\}$ is such that the rates per unit charge are
proportionally fair.  That is, if $\{\hat{x}_i\}$ is any other set of rates
then
\begin{equation}
\sum_i w_i \frac{\hat{x}_i - x_i^*}{x_i^*} \le 0.
\label{eq:propfair}
\end{equation}
Now let $\{\hat{u}_{i,k}\}$ be any set of subflow rates not equal to
those found by (\ref{eq:subflow_alloc}) and let 
$\hat{x}_i = \sum_k \hat{u}_{i,k}$. From (\ref{eq:subflow_alloc}) we have
$w_i / x_i^* = w_{i,k} / u_{i,k}$
for all $i,k$. From (\ref{eq:propfair}) we have
\begin{equation*}
0 \ge \sum_i \frac{w_i}{x_i^*} \sum_{k} (\hat{u}_{i,k} - u_{i,k}) =
\sum_{i} \sum_{k} w_{i,k} \frac{\hat{u}_{i,k} - u_{i,k}}{u_{i,k}}
\end{equation*}
%
%
So the proportionally allocated solution is such that the rates
per unit charge are proportionally fair. Thus it is the unique solution to
(\ref{eq:problem1}) with $f_{i,k}(u_{i,k}) = w_{i,k} \log u_{i,k}$.

\subsubsection{Weighted Power Utilities (Negative Exponent)}

As another example, let 
$f_{i,k}(u_{i,k}) = -w_{i,k} u_{i,k}^{-a}$
with $a \ge 1$. These functions are also part of the $\alpha$-fair family.
When $a=1$, the allocation satisfies minimum potential delay fairness and
as $a \to +\infty$, the allocation is
max-min fair \cite{srikant2013communication}. 
The utilities are also of Legendre type and we can use
Theorem \ref{theorem:legendre_sp}. We have 
$f_{i,k}'(u) = a w_{i,k} u^{-(a+1)}$ and
$g'_{i,k}(v) = ( a w_{i,k} /v )^{\frac{1}{a+1}}$.
Next we have $g'_i(y) = \sum_k g'_{i,k} (y) = f_i'^{-1}(y)$. Thus
$f_i'(x) = (a/x^{a+1}) ( \sum_k w_{i,k}^{\frac{1}{a+1}} )^{a+1}$
and the optimum subflow rates are
\begin{equation*}
u_{i,k}^*  = g'_{i,k} (f_i'(x_i^*)) =
\frac{w_{i,k}^{\frac{1}{a+1}}}  {\sum_{k'} w_{i,k'}^{\frac{1}{a+1}}} x_i^*.
\end{equation*}
The utility functions for problem (\ref{eq:problem2}) are
$f_i(x_i) = -x_i^{-a} ( \sum_k  w_{i,k}^{\frac{1}{a+1}} )^{a+1}$.
Again, all of the above functions have domain $\mathbb{R}_{++}$.

\subsubsection{Quadratic Utilities}

Quadratic functions are not of Legendre type and are not necessarily increasing
on $\mathbb{R}_+$, rendering them unsuitable for use as utility
functions. However, the aggregate flow decomposition can be useful when
implementing a gradient projection algorithm. In a gradient projection
algorithm, steepest ascent iterations are followed by projections onto the
feasible set \cite{bertsekas1999nonlinear}. Such a projection is a quadratic
program (QP) that can be simplified by decomposing with supremal convolutions.

Let $\{z_{i,k}\}$ be the set of variables obtained after an iteration of
steepest ascent for problem (\ref{eq:problem1}). This set must be projected
onto the routing polytope $\{ \{u_{i,k}\} : \sum_i \sum_k R_{l,i} u_{i,k} \le
c_l, l=1, \dots, L\}$.  The projection QP is problem (\ref{eq:problem1}) with
quadratic utility $f_{i,k}(u_{i,k}) = -\frac{1}{2}(u_{i,k} - z_{i,k})^2$, and
domain $\{u_{i,k} \ge 0\}$. (In this section all functions are equal to
$-\infty$ outside their domains). Note that $f_{i,k}$ is not Legendre
type. However $f_{i,k}$ is closed, concave, and proper on $\mathbb{R}$ 
and has conjugate $f_{i,k}^*(y) = -\frac{1}{2}y^2 + z_{i,k} y$, with domain
$\{y \le z_{i,k}\}$. Thus any point less than $z_{i,\min} = \min_k z_{i,k}$
lies in $\text{relint}( \text{dom} f_{i,k}^*)$ for all $(i,k)$, and therefore
(\ref{eq:supremal_conj}) can be used to find $f_i$.
The conjugate aggregate utility is
$f_i^*(y) =\sum_k f_{i,k}^*(y) = -K_i y^2/2 + \bar{z}_i y$, 
with domain $\{y \le z_{i,\min}\}$, 
where $\bar{z}_i = \sum_k z_{i,k}$. The aggregate
function is obtained by conjugating $f_i^*$, which yields
$f_i(x_i) = -\frac{1}{2K_i} (x_i - \bar{z}_i)^2$, with domain
$\{x_i \ge \bar{z}_i - K_i z_{i,\min}\}$.
Finally, for each class $i$, the subflows minimize
$\sum_{k} \frac{1}{2} (u_{i,k} - z_{i,k})^2$ subject to
$\sum_{k} u_{i,k} = x^*_{i}$ and $u_{i,k} \ge 0$ for each $k$,
where $x_i^*$ is the solution to the aggregate problem. 
(Thus $x_i^* \ge \bar{z}_i - K_i z_{i,\min}$.)
The subflow problem is
strictly convex and has unique solution
$u_{i,k}^* = z_{i,k} + \frac{1}{K_i} (x_i^* - \bar{z}_i)$. 

\subsubsection{Piecewise Linear Utilities}

Piecewise linear functions are important as they are often used as
approximations of functions that are difficult to work with analytically or are
incompletely known \cite{rockafellar1984network}.  In this case
Theorem \ref{theorem:legendre_sp} is not applicable but 
supremal convolutions can be calculated using
(\ref{eq:supremal_conj}).  Let $f_{i,k}$ be concave and piecewise linear with
non-negative breakpoints $0 = c_1 < c_2 < \dots < c_B$ and corresponding
non-negative slopes $m_1 > m_2 > \dots > m_B = 0$, and let 
$f_{i,k}(c_1) = f_{i,k}(0) = 0$ and $f_{i,k}(x) = -\infty$ for $x < 0$. 
(The number of breakpoints $B$ need not be the same for all utilities.)
Then $f_{i,k}$ is closed and
from \cite[Sec. 8F]{rockafellar1984network}, the conjugate of
$f_{i,k}$ is also concave and piecewise linear with breakpoints 
$0 = m_B < m_{B-1} < \dots < m_1$ and corresponding slopes 
$c_B > c_{B-1} > \dots > c_1$, and $f_{i,k}^*(m_1) = 0$.  
That is, the breakpoints of $f_{i,k}^*$ are the slopes
of $f_{i,k}$ and the slopes of $f_{i,k}^*$ are the breakpoints of $f_{i,k}$.
Finally, $\text{dom} f_{i,k}^* = \mathbb{R}_+$ and thus (\ref{eq:supremal_conj})
can be used.

The aggregate utility function $f_i$ can be found with the following
prescription: For each $f_{i,k}$, find $f_{i,k}^*$ by exchanging breakpoints
and slopes, as described above. Sum these conjugates to find the conjugate of
the aggregate utility $f_i^* = \sum_k f_{i,k}^*$. Thus $f_i^*$ is piecewise
linear and concave as well. Finally, exchange slopes and breakpoints of $f_i^*$
to arrive at $f_i$.

Therefore, the piecewise-linear problem, which is a linear program (LP)
in $K = \sum_i K_i$ variables, can be decomposed into one LP in $N$
variables, followed by $N$ parallel sub-LP's, the $i$th sub-LP having
$K_i$ variables.

\section{Extension to Multipath Case}
\label{sec:multipath}

Now suppose that for each flow class $i$, traffic may be split into subflows
and routed over multiple paths. (Here, a subflow refers to that portion of a
flow routed over a certain path, as opposed to a constituent flow of a flow
class).  Let $J$ be the number of paths and assume $J$ is the same for all flow
classes. For each class $i$, define the $L \times J$ routing matrix $\bS_i$ as
\begin{equation*}
[\bS_i]_{l,j} = \left\{
\begin{tabular}{ll}
$1$, & 
\begin{tabular}{@{}l@{}}
traffic on the $j$th path of class $i$ \\
passes through link $l$ \
\end{tabular}  \\
$0$, & otherwise.
\end{tabular}
\right.
\end{equation*}
and let the overall $L \times NJ$ routing matrix be 
$\bR = [\bS_1, \dots, \bS_N]$.
Finally, let $u_{i,j,k}$ be the rate on the $j$th path of flow
$k$ of class $i$. The multipath utility maximization problem is
%
\begin{eqnarray}
\underset{\{u_{i,j,k} \}}{\text{maximize}} &~& 
\sum_{i=1}^N \sum_{k=1}^{K_i} f_{i,k} ( \sum_{j=1}^J u_{i,j,k})  
\nonumber\\
\text{subject to} &~& \sum_{i=1}^N \sum_{j=1}^J \sum_{k=1}^{K_i} 
[\bS_i]_{l,j} u_{i,j,k} \le c_l, \ \forall l \nonumber\\
&~& u_{i,j,k} \ge 0, \ \forall i,j,k.
\label{eq:problem1m}
\end{eqnarray}
%
%
Letting $x_{i,j} = \sum_k u_{i,j,k}$ be the aggregate rate on path $j$ of class
$i$, the aggregate flow problem in the multipath case is
%
\begin{eqnarray}
\underset{\{x_{i,j} \}}{\text{maximize}} &~& 
\sum_{i=1}^N f_{i} ( \sum_{j=1}^J x_{i,j} )  \nonumber\\
\text{subject to} &~& 
\sum_{i=1}^N \sum_{j=1}^J [\bS_i]_{l,j} x_{i,j} \le c_l, \
\forall l \nonumber\\
&~& x_{i,j} \ge 0, \ \forall i,j.
\label{eq:problem2m}
\end{eqnarray}
%
For these problems, explicit constraints for non-negativity of the throughputs
are added because, for example $\sum_j u_{i,j,k}$ can be non-negative even
with some negative subflows.
Unlike the single-path case, neither problem is strictly convex. 

\subsection{Multipath Supremal Convolution Decomposition}

Let the $f_{i,k}$'s be concave and proper on $\mathbb{R}$ with 
$\text{dom} f_{i,k} \subset \mathbb{R}_+$. For each $i,k$ pair, define
\begin{equation*}
\phi_{i,k}(\bu_{i,k}) = \left\{
\begin{tabular}{ll}
$f_{i,k}(\b1_J^T\bu_{i,k})$, & $\bu_{i,k} \in \mathbb{R}^J_+$ \\
$-\infty$, & otherwise
\end{tabular}
\right.
\end{equation*}
where $\bu_{i,k} = [u_{i,1,k}, \dots, u_{i,J,k}]^T$. Then $\phi_{i,k}$ is
concave and proper on $\mathbb{R}^J$ (but not strictly concave, even if
$f_{i,k}$ is) and problem (\ref{eq:problem1m}) is equivalent to
\begin{equation}
\underset{\{\bu_{i,k} \}}{\text{maximize}} \quad
\sum_{i=1} \sum_{k=1} \phi_{i,k} ( \bu_{i,k} )
\label{eq:problem1mb}
\end{equation}
subject to the link load constraints of (\ref{eq:problem1m}).
Define the aggregate flow problem by
\begin{equation}
\underset{\{\bx_{i} \}}{\text{maximize}} \quad
\sum_{i=1} \phi_{i} (\bx_i)  
\label{eq:problem2mb}
\end{equation}
with the link load constraints of (\ref{eq:problem2m}).
Here $\phi_i(\bx_i)$ is a function from $\mathbb{R}^J$ to
$\mathbb{R}$ and $\bx_i = [x_{i,1}, \dots, x_{i,J}]^T$.
Similar to the argument in Section \ref{sec:sup_conv_decomp},
problem (\ref{eq:problem2mb}) is equivalent to problem 
(\ref{eq:problem1mb}) with concave aggregate functions  
\mbox{$\phi_i = \phi_{i,1} \diamond \dots \diamond \phi_{i,K_i}$}
if for each flow class $i$, the optimal subflow rates solve the problem
\begin{equation*}
\underset{\{\bu_{i,k} \}}{\text{maximize}} 
\sum_{k=1}^{K_i} \phi_{i,k} (\bu_{i,k}), \quad
\text{subject to} \sum_{k=1}^{K_i} \bu_{i,k} = \bx_i^*
\end{equation*}
where $\bx_i^*$ is the solution of problem (\ref{eq:problem2mb}) (provided it
exists) with $\phi_i$ defined as above.  If the $f_{i,k}$'s are closed and
$\cap_k \text{relint}(\text{dom} f_{i,k}^*) \neq \emptyset$, then it can
be shown that the same
is true of the $\phi_{i,k}$'s and the aggregate functions can be found
using 
$\phi_i = (\sum_k \phi_{i,k}^*)^*$ where 
$\phi^*(\by) = \inf_{\bu} (\bu^T \by - \phi(\bu))$.

\subsection{Multipath Legendre-Type Case}

As in the single-path case, when $\text{dom} f_{i,k} = \mathbb{R}_{++}$, strong
duality of (\ref{eq:problem1m}) and (\ref{eq:problem2m}) follows from Slater's
condition. However, in the multipath case, neither problem is strictly convex
and uniqueness of the solutions cannot be guaranteed.

\begin{theorems}
Let $f_{i,k}$, $g_{i,k}$, $f_i$, and $g_i$ satisfy the conditions
of Theorem \ref{theorem:legendre_sp} for all $i,k$.
Let $\{x_{i,j}^*\}$ be a primal solution to
(\ref{eq:problem2m}) with this definition of $\{f_i\}$. Then a 
solution of the following constrained system of linear equations
\begin{eqnarray}
\sum_k u_{i,j,k} &=&  x_{i,j}^*, \ \forall i,j \nonumber\\
\label{eq:mp_usolution}
\sum_j u_{i,j,k} &=&  g_{i,k}' (f_i'(\bar{x}_i^*)), \ \forall i,k \\
u_{i,j,k} &\ge& 0, \ \forall i,j,k \nonumber
\end{eqnarray}
(where $\bar{x}_i^* = \sum_j x_{i,j}^*$), is a primal solution to
(\ref{eq:problem1m}).
Furthermore, if $\blambda^*$ is the dual solution to (\ref{eq:problem2m})
corresponding to the link load constraints and $\mu_{i,j}^*$ is 
the dual solution to (\ref{eq:problem2m}) corresponding to the 
non-negativitiy constraint of $x_{i,j}$, then the dual solution to
(\ref{eq:problem1m}) corresponding to the link load constraints is
$\brho^* = \blambda^*$ and the dual solution to (\ref{eq:problem1m}) 
corresponding to the non-negativity constraint of $u_{i,j,k}$ is
$\sigma_{i,j,k}^* = \mu_{i,j}^*$, for each $k$.
\end{theorems}
%
%

\begin{table*}[!t]
\footnotesize
\caption{Comparison of ADMM, CP, and Grad. Proj. for Small Graph Example}
\centering
\begin{tabular}{r|rrrr|rrrr|rrrr}
\hline\hline
~ & \multicolumn{4}{|c|}{ADMM} & \multicolumn{4}{c|}{Gradient Projection} &
\multicolumn{4}{c}{Chambolle-Pock} \\
\multicolumn{1}{c}{$N$} & 
\multicolumn{1}{|c}{$f^*$} & 
\multicolumn{1}{c}{$l_{\max}$} & 
\multicolumn{1}{c}{$n_{\text{iter}}$} &
\multicolumn{1}{c}{$t$ (sec)} & 
\multicolumn{1}{|c}{$f^*$} & 
\multicolumn{1}{c}{$l_{\max}$} & 
\multicolumn{1}{c}{$n_{\text{iter}}$} & 
\multicolumn{1}{c}{$t$ (sec)} &
\multicolumn{1}{|c}{$f^*$} & 
\multicolumn{1}{c}{$l_{\max}$} & 
\multicolumn{1}{c}{$n_{\text{iter}}$} & 
\multicolumn{1}{c}{$t$ (sec)} \\
\hline
10 & -92.084 & 10.000 & 194 & 0.0193 & -92.085 & 10.000 & 431 & 1.1607
& -92.084 & 10.000 & 74 & 0.0030 \\
15 & -136.800 & 10.000 & 207 & 0.0259 & -136.809 & 10.000 & 500 & 1.2534 
& -136.800 & 10.000 & 112 & 0.0050\\
20 & -182.002 & 10.000 & 304 & 0.0446 & -182.002 & 10.000 & 590 & 1.6405 
& -182.002 & 10.000 & 258 & 0.0130\\
25 & -243.806 & 10.000 & 296 & 0.0508 & -243.806 & 10.000 & 1288 & 3.4499 
& -243.806 & 10.000 & 225 & 0.0130 \\
30 & -289.040 & 10.000 & 296 & 0.0574 & -289.040 & 10.000 & 2029 & 5.9341 
& -289.040 & 10.000 & 256 & 0.0170 \\
\end{tabular}
\label{table:small_example}  
\end{table*}

{\it Proof.}
Let $h_{i,k} = g'_{i,k} = f'^{-1}_{i,k}$ for each $i,k$ and let
$\bar{u}_{i,k} = \sum_j u_{i,j,k}$.
The Lagrangian for problem (\ref{eq:problem1m}) is
\begin{alignat*}{2}
\calL_1(\bu, \brho, \bsigma) &=&& \sum_{i,k} f_{i,k} (\bar{u}_{i,k}) - 
\sum_{l=1}^L \rho_l ( \sum_{i,j,k} [\bS_i]_{l,j} u_{i,j,k} - c_l) \\
&~&& + \sum_{i,j,k} \sigma_{i,j,k} u_{i,j,k}.
\end{alignat*}
Setting the derivative with respect to $u_{i,j,k}$ to zero gives
\begin{equation}
f_{i,k}'(\bar{u}_{i,k}) = [\bS_i^T \brho]_j - \sigma_{i,j,k}, \ \forall i,j,k.  
\label{eq:mp_lagrangian_deriv}
\end{equation}
Thus, the following seven relations constitute the KKT conditions for
problem (\ref{eq:problem1m}):
\begin{alignat}{2}
\label{eq:kkt1.1m}
\sum_{i,j,k} [\bS_i]_{l,j} u_{i,j,k} \ &\le&& \ c_l, \ \forall l \\
\label{eq:kkt1.2m}
u_{i,j,k} &\ge&& \ 0, \ \forall i,j,k \\
\label{eq:kkt1.3m}
\brho \ &\ge&& \ \mathbf{0} \\
\label{eq:kkt1.4m}
\bsigma \ &\ge&& \ \mathbf{0} \\
\label{eq:kkt1.5m}  
\rho_l ( \sum_{i,j,k} [\bS_i]_{l,j} u_{i,j,k} - c_l ) \ &=&& \ 0, 
\ \forall l \\
\label{eq:kkt1.6m}  
\sigma_{i,j,k} u_{i,j,k} &=&& \ 0, \ \forall i,j,k \\
\label{eq:kkt1.7m}
h_{i,k} ([\bS_i^T \brho]_j - \sigma_{i,j,k})
\ &=&& \ 
\bar{u}_{i,k},
\ \forall i,j,k.
\end{alignat}
Next, turning to problem (\ref{eq:problem2m}) with $f_i=g_i^*$, 
let $h_i = \sum_k h_{i,k} = \sum_k g'_{i,k} = g_i'$ 
for each $i$ and let $\bar{x}_i = \sum_j x_{i,j}$. The Lagrangian is
%
\begin{alignat*}{2}
\calL_2(\bx, \blambda, \bmu) &=&& \sum_{i} f_{i} (\bar{x}_{i}) - 
\sum_{l=1}^L \lambda_l ( \sum_{i,j} [\bS_i]_{l,j} x_{i,j} - c_l ) \\
&~&& + \sum_{i,j} \mu_{i,j} x_{i,j}.
\end{alignat*}
The KKT conditions for problem (\ref{eq:problem2m}) are
\begin{eqnarray*}
\sum_{i,j} [\bS_i]_{l,j} x_{i,j} &\le& c_l, \ \forall l \\
x_{i,j} &\ge& 0, \ \forall i,j \\
\blambda &\ge& \mathbf{0} \\
\bmu &\ge& \mathbf{0} \\
\lambda_l ( \sum_{i,j} [\bS_i]_{l,j} x_{i,j} - c_l ) &=& 0,
\ \forall l\\
\mu_{i,j} x_{i,j} &=& 0, \ \forall i,j\\
\bar{x}_i &=& h_i([\bS_i^T \blambda]_j - \mu_{i,j}), \ \forall i,j
\end{eqnarray*}
Now set $\brho = \blambda^*$ and $\sigma_{i,j,k} = \mu_{i,j}^*$ for all
$i,j,k$ and let $\{u_{i,j,k}^*\}$ be a solution to 
(\ref{eq:mp_usolution}). Then it can
be seen that all KKT conditions (\ref{eq:kkt1.1m})--(\ref{eq:kkt1.7m}) are
satisfied, and furthermore, using (\ref{eq:mp_lagrangian_deriv})
\begin{equation*}
\sum_{j,k} u_{i,j,k}^* = \sum_k h_{i,k}(f_i'(\bar{x}_i^*)) = 
h_i(f_i'(\bar{x}_i^*)) = \bar{x}_i^*
\end{equation*}
which ensures $\sum_{j,k} u_{i,j,k}^* = \sum_j x_{i,j}^*$. \qed

\begin{figure}[!t]
\centering
\includegraphics[width=1.25in]{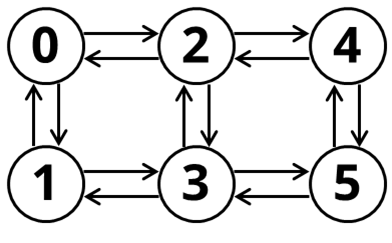}
\caption{Small example graph}
\label{fig:small_graph}
\end{figure}

\begin{figure}[!t]
\centering
\includegraphics[width=2.5in]{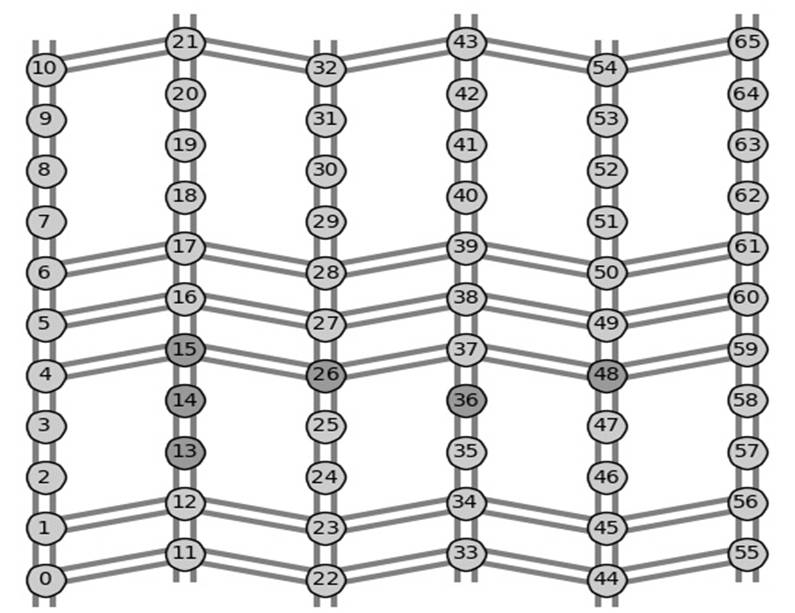}
\caption{Large example graph}
\label{fig:irid_net}
\end{figure}

Note that when $J=1$, the second equation of problem
(\ref{eq:mp_usolution}) reduces to (\ref{eq:sp_usolution}) which 
guarantees that the first equation and the non-negativity condition hold.

The subflow allocation problem given by 
(\ref{eq:mp_usolution}) can be decomposed
into $N$ parallel problems (one for each class $i$). 
Let $\bu_{i,k} = [u_{i,1,k}, \dots, u_{i,J,k}]^T$,
$\bu_i = [\bu_{i,1}^T, \dots, \bu_{i,K_i}^T]^T$, 
$\bx_i = [x_{i,1}, \dots, x_{i,J}]^T$, and define the matrices
$\bA_i = \b1_{K_i}^T \otimes \bI_J$ and $\bB_i = \bI_{K_i} \otimes \b1_J^T$.
Then the $i$th optimal subflow vector $\bu_i^*$ solves
\begin{eqnarray}
\label{eq:mp_u_lp}
\left[ 
  \begin{tabular}{c}
    $\bA_i$ \\
    $\bB_i$
  \end{tabular}
\right] \bu_i 
&=& 
\left[
  \begin{tabular}{c}
    $\bx_i^*$ \\
    $\bg_i$
  \end{tabular}
\right]
 \\
\bu_i & \ge & \bf0 \nonumber
\end{eqnarray}
where $\bg_i = [g_{i,1}' (f_i'(\bar{x}_i^*)), \dots, g_{i,K_i}' (f_i'(\bar{x}_i^*))]^T$. 

Note that the only component of (\ref{eq:mp_u_lp}) that depends on
the utility functions is $\bg_i$. As an example, for the case of weighted
logarithm utilities (see Section \ref{sec:weighted_log_example}) with
$f_{i,k}(u) = w_{i,k} \log u$, we have $\bg_i = (\bar{x}_i^* / \bar{w}_i) \bw_i$,
where $\bw_i = [w_{i,1}, \dots, w_{i,K_i}]^T$, and 
$\bar{w}_i = \sum_k w_{i,k}$.

\section{Aggregate Decomposition with ADMM}
\label{sec:algorithms}

Here we show that the alternating direction method of multipliers (ADMM)
algorithm \cite{boyd2011distributed} can
inherently decompose problem (\ref{eq:problem1}) into an optimization over
aggregate flows and $N$ parallel optimizations over the constituent flows.
Assume that the utility functions $f_{i,k}$ have domain $\mathbb{R}_{+}$. 

\subsection{ADMM Algorithm}

To apply ADMM to problem (\ref{eq:problem1}) we recast it as
\begin{eqnarray}
&\underset{\{u_{i,k} \in \mathbb{R}_+, x_i \in \mathbb{R}, 
y_l \in \mathbb{R}\}}
{\text{minimize}} & 
\sum_i \sum_{k} -f_{i,k} (u_{i,k}) + h(\by)  \nonumber\\
&\text{subject to} & \sum_{k} u_{i,k} =x_i, \
i = 1, \dots, N \nonumber\\
&~&\by = \bR \bx
\label{eq:prob_admm}
\end{eqnarray}
where $\bx = [x_1, \dots, x_N]^T$ and $\by = [y_1, \dots, y_L]^T$.
The function $h$ indicates that the links
are not overloaded. That is $h(\by) = 0$ if $\by \le \bfc$, and $+\infty$
otherwise, with
$\bfc = [c_1, \dots, c_L]^T$.
The augmented Lagrangian for problem (\ref{eq:prob_admm}) is 
\begin{alignat*}{1}
\calL_r (\bu, & \bx, \by, \blambda, \brho)  =
\sum_i \sum_k -f_{i,k}(u_{i,k}) + h(\by) + \blambda^T(\bs - \bx) \\
& + \brho^T(\by - \bR \bx) 
+ \frac{r}{2}( \|\bx - \bs \|^2 + \| \bR \bx - \by \|^2)
\end{alignat*}
where $\bs = \left[ \sum_k u_{1,k}, \dots, \sum_k u_{N,k} \right]^T$.
%
Here $\blambda \in \mathbb{R}^N$ and $\brho \in \mathbb{R}^L$ are the dual
variables and $r$ is the penalty parameter.  The ADMM method involves repeated
minimizations of $\calL_r$ over $(\bu, \by)$, and then $\bx$. 
The minimizer with respect to $\bx$ is $\bx = \bA^{-1} \bfb$ 
%
where $\bA = \bI + \bR^T \bR$ and 
$\bfb = \bs + \bR^T \by + (\blambda + \bR^T \brho) / r$.
Minimization of $\calL_r$ with respect to $\by$ is decoupled from
that of $\bu$ and is achieved by
a simple projection of $\bR \bx - \brho/r$ 
onto the box $\{ \by: \by \le \bfc \}$.
Finally, minimization with respect to $\bu$ involves $N$ 
parallel minimizations of the form
\begin{alignat}{1}
\underset{\{u_{i,k} \in \mathbb{R}_+\}}{\text{minimize}} \  &
\sum_k -f_{i,k}(u_{i,k}) + \lambda_i ( \sum_k u_{i,k} - x_i ) +
\nonumber\\
& \frac{r}{2}  ( \sum_k u_{i,k} - x_i )^2.
\label{eq:admm_opt_u}
\end{alignat}

\subsection{Numerical Examples}

Here we apply the ADMM algorithm with the aggregate flow decomposition to a few
example cases and compare performance against a gradient projection algorithm
and the primal-dual algorithm of Chambolle and Pock \cite{chambolle2011first}.
We examine two sample graphs. The first, which we call the {\it small graph},
shown in Figure \ref{fig:small_graph}, has $M=6$ nodes and $L=14$ links. The
maximum number of source-destination pairs is $N_{\max} = 30$. The second
example graph, the {\it large graph} is shown in
Figure \ref{fig:irid_net}. This represents the topology of the Iridium low
earth orbit satellite constellation
\cite{pratt1999operational}, and includes
$M=66$ satellites (nodes) and $L=192$ links (the actual topology changes as
satellites enter and exit polar regions).  The dark nodes in
Figure \ref{fig:irid_net} represent satellites linked to ground stations which
connect flows to terrestrial networks. Thus, we assume all flows either
originate or terminate at one of these nodes. The resulting maximum number of
source-destination pairs is $N_{\max}=750$.  For both example graphs, we set
all link capacities to 10 units. Thus $\bfc = 10 \cdot \b1_L$.  In all
examples, the number of flows in any flow class (source-destination pair) is
uniformly distributed between 10 and 20. The total number of flows is thus
$15N$ on average. Finally, the path (route) for each source-destination pair
is found using Dijkstra's algorithm.

\subsubsection{ADMM with Weighted Logarithm Utilities}

We let the utility function for the $k$th flow of class $i$ be
$f_{i,k}(u_{i,k}) = w_{i,k} \log u_{i,k}$ with weights $w_{i,k}$ chosen
uniformly from $(0,1)$ and solve the optimization problem with ADMM.
The minimizer of the augmented Lagrangian with 
respect to the individual flows $\{u_{i,k}\}$ is
found by solving (\ref{eq:admm_opt_u}) for each $i$ which gives
\begin{equation*}
u_{i,k}^* = \frac{2 w_{i,k}}{ \psi_i + \sqrt{\psi_i^2 + 4 r \bar{w}_i}}  > 0,  
\end{equation*}
%
where $\psi_i = \lambda_i - rx_i$ and $\bar{w}_i = \sum_k w_{i,k}$.
The ADMM iteration is then given by
\begin{alignat*}{2}
\psi_i^{(n+1)} &=&& \lambda_i^{(n)} - r x_i^{(n)} \\
\bu_i^{(n+1)} &=&& \ 2\bw_i  [\psi_i^{(n+1)} + 
((\psi_i^{(n+1)})^2 + 4r \bar{w}_i)^{1/2} ]^{-1} \\
\by^{(n+1)} &=&& \ [\bR \bx^{(n)} - \brho^{(n)} / r]^+ \\
\bx^{(n+1)} &=&& \ \bA^{-1} (\bs^{(n+1)} + \blambda^{(n)} / r  + 
\bR^T (\by^{(n+1)} + \brho^{(n)} / r)) \\
\blambda^{(n+1)} &=&& \ \blambda^{(n)} + r(\bs^{(n+1)} - \bx^{(n+1)}) \\
\brho^{(n+1)} &=&& \ \brho^{(n)} + r(\by^{(n+1)} - \bR \bx^{(n+1)})
\end{alignat*}
%
where 
$[ \cdot ]^+$ represents projection onto the box $\{ \by: \by \le \bfc \}$,
and
$\bs^{(n+1)} = [ \sum_k u_{1,k}^{(n+1)}, \dots, \sum_k u_{N,k}^{(n+1)} ]^T$.

\subsubsection{Gradient Projection Optimizer}

We compare the ADMM algorithm with a simple gradient projection optimizer.
The gradient projection optimizer utilizes Theorem \ref{theorem:legendre_sp}
with aggregate utilities $f_i(x_i) = \bar{w}_i \log x_i$ and optimal subflow
rates $u_{i,k}^* = w_{i,k} x_i^* / \bar{w}_i$ (see Section 
\ref{sec:weighted_log_example}). The update rule for the aggregate problem
is 
\begin{eqnarray*}
\nabla f(\bx^{(n)})_i &=&  \bar{w}_i / x_i^{(n)} \\
\bx^{(n+1)} &=& P_{\bR}(\bx^{(n)} + \alpha \nabla f(\bx^{(n)}))
\end{eqnarray*}
where $\alpha > 0$ is a step size and $P_{\bR}$ is the function which 
projects onto
the routing polytope $\{\bx : \bR \bx \le \bfc, \bx \ge \bf0\}$. In all
examples that follow, $P_{\bR}$ , which solves a QP, is implemented using the
CVXOPT QP solver \cite{ADV:15}.

\begin{table}[!t]
\footnotesize
\caption{Parameters for Small Graph Example}
\centering
\begin{tabular}{r|rr|r|rrr}
\hline\hline
~ & \multicolumn{2}{|c|}{ADMM} & \multicolumn{1}{c|}{Grad Proj} &
\multicolumn{3}{c}{Chambolle-Pock} \\
\multicolumn{1}{c}{$N$} & 
\multicolumn{1}{|c}{$r$} & 
\multicolumn{1}{c}{pct} & 
\multicolumn{1}{|c}{$\alpha$} & 
\multicolumn{1}{|c}{$\sigma$} & 
\multicolumn{1}{c}{$\tau$} & 
\multicolumn{1}{c}{$\theta$} \\
\hline
10 & 20 & $10^{-4}$ & $1.07 \times 10^{-2}$ & 1.0 & 0.020 & 1.0 \\
15 & 20 & $10^{-4}$ & $1.07 \times 10^{-2}$ & 1.0 & 0.015 & 1.0 \\
20 & 20 & $10^{-4}$ & $1.31 \times 10^{-2}$ & 1.0 & 0.015 & 1.0 \\
25 & 20 & $10^{-4}$ & $6.17 \times 10^{-3}$ & 1.0 & 0.015 & 1.0 \\
30 & 20 & $10^{-4}$ & $6.19 \times 10^{-3}$ & 1.0 & 0.013 & 1.0 
\end{tabular}
\label{table:small_example_prm}  
\end{table}

\begin{table*}[!t]
\footnotesize
\caption{Comparison of ADMM and CP for Large Graph Example}
\centering
\begin{tabular}{r|rrrrrr|rrrrrrr}
\hline\hline
~ & \multicolumn{6}{|c|}{ADMM} & \multicolumn{7}{c}{Chambolle-Pock} \\
\multicolumn{1}{c}{$N$} & 
\multicolumn{1}{|c}{$r$} & 
\multicolumn{1}{c}{pct} & 
\multicolumn{1}{c}{$f^*$} & 
\multicolumn{1}{c}{$l_{\max}$} & 
\multicolumn{1}{c}{$n_{\text{iter}}$} &
\multicolumn{1}{c}{$t$ (sec)} & 
\multicolumn{1}{|c}{$\sigma$} & 
\multicolumn{1}{c}{$\tau$} & 
\multicolumn{1}{c}{$\theta$} & 
\multicolumn{1}{c}{$f^*$} & 
\multicolumn{1}{c}{$l_{\max}$} & 
\multicolumn{1}{c}{$n_{\text{iter}}$} & 
\multicolumn{1}{c}{$t$ (sec)} \\
\hline
%
%
50 & 40 & $10^{-4}$ & -1326.781 & 10.000 & 100 & 0.0358 & 
10.0 & $3.0 \times 10^{-4}$ & 1.0 & -1326.780 & 10.000& 70 & 0.0120 \\
75 & 40 & $10^{-4}$ & -2002.522 & 10.001 & 162 & 0.0777 & 
10.0 & $2.0 \times 10^{-4}$ & 1.0 & -2002.522 & 10.000 & 314 & 0.0760 \\
100 & 40 & $10^{-4}$ & -2589.978 & 10.000 & 179 & 0.1093 & 
10.0 & $5.0 \times 10^{-4}$ & 0.1 & -2589.978 & 10.000 & 745 & 0.2280 \\
125 & 40 & $10^{-4}$ & -3333.174 & 10.007 & 208 & 0.1529 &
10.0 & $4.9 \times 10^{-4}$ & 0.1 & -3333.174 & 10.000 & 843 & 0.3119 \\
\end{tabular}
\label{table:large_example}  
\end{table*}

\subsubsection{Chambolle-Pock Optimizer}

Problem (\ref{eq:problem1}) can be solved with the Chambolle-Pock (CP) algorithm
by writing it as
\begin{equation*}
\underset{\{\bu \in \mathbb{R}^K_{++}\}}{\text{minimize}} 
\quad -\sum_i \sum_k f_{i,k} (u_{i,k}) + g(\bQ \bu)
\end{equation*}
where $\bu$ is the concatenation of the $N$ subflow rate vectors $\{\bu_i\}$
and $g$ is the indicator function of the box  $\{ \by: \by \le \bfc \}$.
The $L \times K$ matrix $\bQ$ is defined by
\begin{equation*}
\bQ = [ \underbrace{\bfr_1, \dots, \bfr_1}_{K_1}, \dots, 
\underbrace{\bfr_N, \dots, \bfr_N}_{K_N} ]
\end{equation*}
The algorithm requires evaluation of the proximal operators 
\cite{rockafellar1970convex} of $\sigma g^*$ and $\tau f$ where
$f(\bu) = -\sum_{i,k} f_{i,k}(u_{i,k})$, 
$g^*$ is the convex conjugate of $g$, and $\sigma$ and $\tau$ are positive
constants. Using Moreau's theorem
\cite[Theorem 31.5]{rockafellar1970convex} we get
$\prox_{\sigma g^*}(\bz) = \bz - \sigma [\bz / \sigma]^+$
(again $[ \cdot ]^+$ signifies projection onto  $\{ \by: \by \le \bfc \}$).
The proximal operator of $\tau f$ is
\begin{equation*}
\prox_{\tau f}(\bz)_{i,k} = \frac{z_{i,k} + \sqrt{z_{i,k}^2 + 4 \tau w_{i,k}}}{2}.
\end{equation*}
%
%
The algorithm consists of the following iteration
\begin{eqnarray*}
\by^{(n+1)} &=& \prox_{\sigma g^*} (\by^{(n)} + \sigma \bQ \bv^{(n)}) \\
\bu^{(n+1)} &=& \prox_{\tau f} (\bu^{(n)} - \tau \bQ^T \by^{(n+1)}) \\
\bv^{(n+1)} &=& \bu^{(n+1)} + \theta (\bu^{(n+1)} - \bu^{(n)})
\end{eqnarray*}
where $\theta \in [0,1]$.


\subsubsection{Algorithm Comparison with Small Graph}

Each iteration of ADMM contains three sparse (0-1 matrix)-vector multiplies
with $\bR$ and $\bR^T$ and one $N \times N$ set of linear equations with the 
same coefficient matrix $\bA$.
The CP iterations contain two multiplications
with the sparse 0-1 matrices $\bQ$ and $\bQ^T$. Finally, each iteration
of the gradient projection algorithm solves a QP 
(with sparse constraint matrix $\bG = [\bR^T, -\bI_N]^T$). Thus, the
gradient algorithm has the highest per-iteration cost, followed by ADMM and
CP.
 
\begin{figure}[!t]
\centering
\includegraphics[width=3.00in]{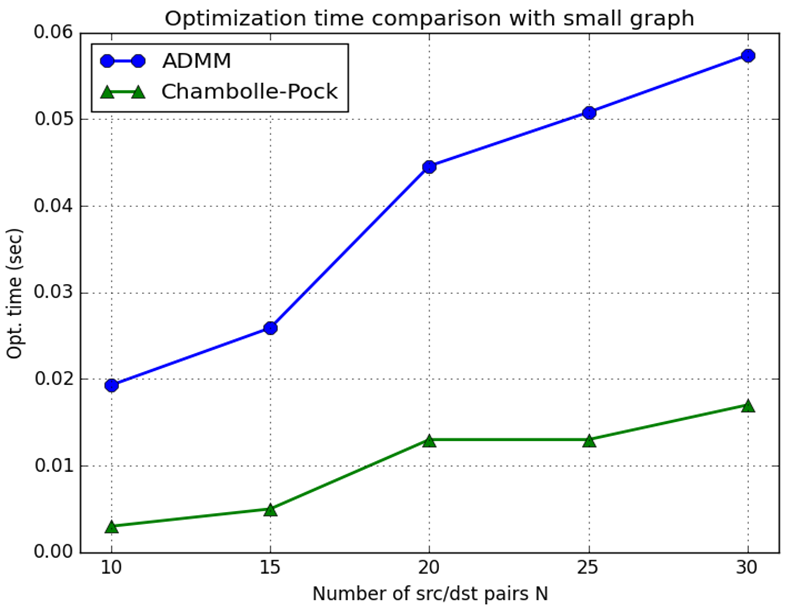}
\caption{Optimization time comparison with small graph example}
\label{fig:opt_time_small}
\end{figure}

The algorithms' performances are summarized in Table \ref{table:small_example}
for various numbers of source-destination pairs $N$. For each algorithm, the
converged objective value $f^*$ is shown, along with the maximum link load
$l_{\max}$, the number of iterations $n_{\text{iter}}$, and the optimization
time $t$. In Table \ref{table:small_example_prm} the algorithm parameters
are listed, including the ADMM penalty 
parameter $r$, the gradient
projection step-size $\alpha$, and the percent threshold (pct). This value is
used as the stopping criterion for ADMM (i.e., when the augmented Lagrangian
changes by less than pct percent, stop). Also shown are the three CP
parameters $\sigma$, $\tau$, and $\theta$.
The gradient projection step-sizes and CP parameters are
individually tuned for fastest convergence.  The optimization times are
averaged over 10 runs (with identical random number generator seeds).  All 
simulations were performed using
Python/Numpy, and the projection step in the gradient projection algorithm
uses the CVXOPT QP solver (which in turn uses the CHOLMOD sparse Cholesky
solver).  The optimization time of ADMM and CP is plotted versus $N$ in Figure
\ref{fig:opt_time_small}. 
From Table \ref{table:small_example}, the number of CP
iterations required for this example is consistently less than the number
of ADMM iterations. As it has a lower per-iteration cost, the convergence
time of CP is lower. The gradient projection algorithm has
the highest per-iteration cost as well as the largest number of iterations, 
and thus converges slowest. Note that, without the aggregate flow decomposition
(Theorem \ref{theorem:legendre_sp}), the gradient projection optimizer would be
far slower. 

\subsubsection{Algorithm Comparison with Large Graph}

\begin{figure}[!t]
\centering
\includegraphics[width=3.00in]{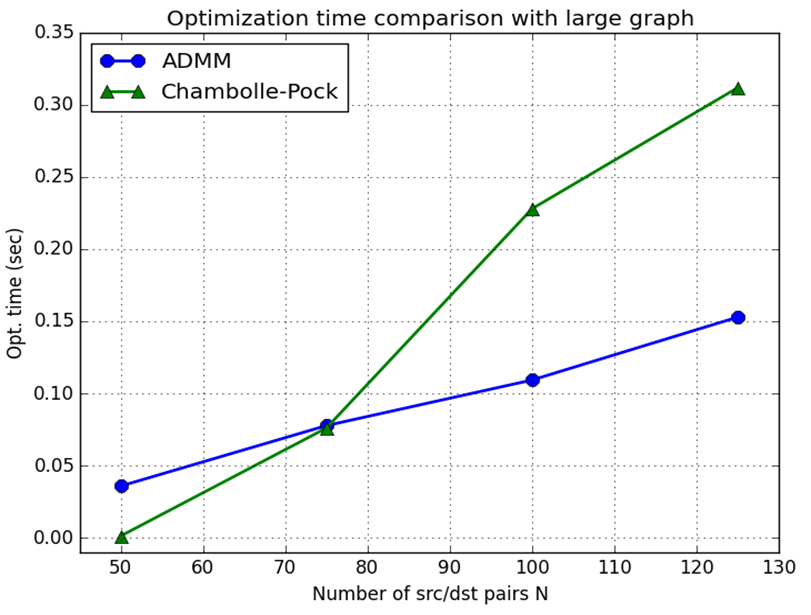}
\caption{Optimization time comparison with large graph example}
\label{fig:opt_time_large}
\end{figure}

Next, we repeat the experiment using the large graph. Table 
\ref{table:large_example} shows the results 
along with the algorithm parameters.
The gradient projection algorithm has been omitted as its convergence
times are far greater than ADMM and CP.
The optimization times for the ADMM and CP algorithms are plotted in 
Figure \ref{fig:opt_time_large}.  In this example, as $N$ increases, the
number of ADMM iterations grows slower than the number of CP
iterations. Thus, although the CP per-iteration cost is lower, the
larger number of iterations for large $N$ renders CP slower than ADMM.


\section{Conclusion}
\label{sec:conclusion}

We have shown that for many types of utilities, the solution to a  
$K$-flow NUM problem can be found by solving a simpler $N$-variable
problem. This principle holds for both single-path and multipath NUM 
problems. The results of this paper have applicability for software-defined
networks in which a controller must solve the global NUM problem.
These results can also be beneficial for networks consisting of several
hub-spoke clusters. For example, with $N$ clusters and $K_i$ sources in the
$i$th cluster, the problem (\ref{eq:problem2}) can be substituted for
problem (\ref{eq:problem1}). This simpler problem could then be solved
in a distributed manner by the hub nodes, which would in turn allocate
subflow rates to the spoke nodes.

\section{Acknowledgements}

The first author wishes to thank Dr. Joseph Yadegar from UtopiaCompression
Corporation and Dr. You Lu of Google for insightful conversations regarding
this work. This research was partly supported by the United States Air Force
under contract number FA9453-14-C-0060. The views and conclusions contained
herein are those of the authors and should not be interpreted as necessarily
representing the official policies or endorsements, either expressed or
implied, of the United States Air Force.

\bibliography{IEEEabrv,robby}
\bibliographystyle{IEEEtran}

\end{document}